\documentclass[aps,twocolumn,superscriptaddress]{revtex4}
\usepackage{amsmath,amssymb}
\usepackage{graphics,graphicx,subfigure}
\usepackage{ulem}
\usepackage{dcolumn,bm}
\usepackage{psfrag}
\usepackage{color}
\usepackage{multirow}
\usepackage{braket}
\newcommand{\la}{\langle}
\newcommand{\ra}{\rangle}
\newcommand{\epsi}{\epsilon}
\newcommand{\cu}
{\affiliation{Department of Physics, University of Calcutta,
92 Acharya Prafulla Chandra Road, Kolkata 700009, India.}}
\newcommand{\be}
{\begin{equation}}
\newcommand{\ee}
{\end{equation}}

\begin{document}

\title{
The variational method applied to the harmonic oscillator in presence of  a delta function
potential}
\author{Indrajit Ghose}
\cu
\author{Parongama Sen}
\cu


\begin{abstract}

The problem of the harmonic oscillator with a centrally located delta function potential can be exactly solved in one dimension where the eigenfunctions are expressed as   superpositions of the Hermite polynomials or 
as  confluent hypergeometric functions in general. The eigenfunctions obtained exactly are difficult to
visualise and  hence to gain more insight, one can  
attempt using   model wave functions which are explicitly and simply expressed.  Here we apply the 
variational method to verify how close one can 
approach the exact ground state eigenvalues using such trial wave functions. 
We obtain the estimates of the ground state energies which are closer to the 
exact values in comparison to earlier approximate  results for both the repulsive and attractive delta potentials. 

\end{abstract}

\maketitle

\section {Introduction}
The harmonic oscillator with the Hamiltonian, 
\begin{equation}
\label{eq:eqn0}
H = -\frac{\hbar ^2}{2m}\nabla^2 + \frac{1}{2}m\omega ^2 r ^2 
\end{equation}
is one of the most well known  Hamiltonians in quantum mechanics  that has been 
exactly solved in one and higher
dimensions, the eigenfunctions here are expressed  as product of a Gaussian function and the  Hermite polynomials. 
The energy eigenvalues are $E_n = (n+1/2) \hbar \omega$ 
with $n = 0, 1, 2, .....$ in one dimension. 
The solutions are of definite parity in all dimensions. 

In this article we consider the harmonic oscillator (HO henceforth) in one dimension in presence of a centrally
located delta function potential such that the Schr\"{o}dinger equation obeyed by the eigenfunctions $\psi$ is 
\begin{equation}
\label{eq:hodeleqn}
-\frac{\hbar ^2}{2m}\frac{d ^2 \psi}{d x^2} + \frac{1}{2}m\omega ^2 x ^2 \psi +\gamma \delta(x) \psi = E\psi .
\end{equation}
Here $\gamma$ is the strength of the potential which is positive (negative) for a repulsive (attractive) potential. This problem  can also be exactly solved. In presence of the delta function, the  odd parity eigenfunctions are not affected. The energy for the even parity eigenfunctions can be obtained from a transcendental equation. 

The  delta function potential demands that there be a discontinuity in the first derivative of $\psi$ at the origin. Problems in quantum mechanics in presence of a delta function potential can be solved  using a general prescription given in \cite{atkinson} in one  or higher dimensions.  
For the harmonic oscillator,  the  solution in Cartesian coordinates can be  expressed   as a superposition of  the eigenstates of the HO, i.e., in terms of the Hermite polynomials. 

Various methods have been used to obtain the eigenvalues  for the harmonic oscillator in a delta function potential exactly \cite{atkinson,avakian,viana-gomes,ferkous} or approximately \cite{patil}. 
In \cite{viana-gomes}, instead of taking the superposition of the 
eigenstates of the free HO,  a  different approach is used from which a lot of insight can be  gained. The eigenfunctions are exactly identified in \cite{viana-gomes}, however, these functions are not obtained in closed form and expressed implicitly in terms of integrals and are therefore not easy to visualise. An important question is therefore whether one can gain useful information by considering model wave functions which have explicit expressions and this approach was taken in \cite{patil}. The expectation value of the energy was calculated using a model wave function and compared with the exact values.

 Variational method is a well known approximation method \cite{griffiths} used to estimate the upper bound of the ground state energy and in some cases higher level energies as well. 
Here  a    trial wave function $\phi$ is used which involves one or more parameters and their optimal values are obtained using the condition that $\langle \phi |H|\phi \rangle$ is a minimum for these values.
Variational  method always overestimates the ground state energy since for any arbitrary trial wave function $\phi$, $\langle \phi |H|\phi \rangle \geq E_0$  where $E_0$ is the actual ground state energy.
 
In this article,  we construct trial wave functions for the HO with a $\delta$ function potential, which obey some essential criteria (boundary conditions etc.) and use the  variational method  to  obtain the optimal values of the parameters. 
 Our aim is to  see how close the actual ground state energy can be approached.    
We have  used one parameter trial wave functions and obtained the results for the ground state energy which  are in fact closer to the exact values compared to those in \cite{patil}, where some model wave functions were used. 
 We also discuss some extreme limits and basic features of the system using the present results.

In section \ref{exact},  we review  the exact result briefly. In sections \ref{sec:attract} and \ref{sec:repulse}, we  present the results for the estimated ground state energy for the attractive and repulsive $\delta$ potential respectively.  In \ref{sec:discuss}, the results are discussed and compared with the existing ones. 

\section{Exact result}
\label{exact}

 In this section, we present  the approach used in \cite{viana-gomes}. The general solutions for the free harmonic oscillator (i.e., without the delta potential) can be expressed in terms of the confluent hypergeometric functions. The energy is written  as  $(\nu + 1/2) \hbar \omega$ and  non-integer $\nu$ values are not allowed  as that leads to some physical inconsistencies, discussed in detail in this section.  For $\nu = 0$ or a positive integer, the solutions are the  well-known Hermite polynomials. 

Changing the variables in equation \eqref{eq:hodeleqn} to dimensionless variables with $y=\frac{x}{a} ,~ a = \sqrt{\frac{\hbar}{m\omega}}$, $\gamma \delta (x)=\frac{\gamma}{a}\delta (y)$, $\epsilon = \frac{ma^2}{\hbar ^2}E = \frac{E}{\hbar \omega}, ~g = \frac{ma\gamma }{\hbar ^2}$, one gets the  reduced Schr\"{o}dinger equation

\be
\frac{d ^2 \psi}{d y^2} + (2\epsilon - y^2)\psi - 2g\delta (y)\psi=0 \label{final_eqn}.
\ee
One can further use the notation $\epsilon = \nu + 1/2$ as in \cite{viana-gomes}. The solution for $g=0$ can be  obtained in the form
\begin{equation}
\label{solution}
\psi (y) = e^{ -\frac{1}{2}y^2}w(y).
\end{equation}
The  solutions for the differential equation obeyed by $w(y)$  are expressed in terms of confluent hypergeometric functions known as Tricomi and Kummer functions \cite{arfken}. 
Non-integer values of $\nu$ are  not acceptable for $g=0$ for the following reasons:   the Tricomi functions have a discontinuous derivative at the origin for non-integer $\nu$ and the Kummer functions, though  smooth at the origin,   blow up at infinity. 
However, these problems disappear for  integer values of $\nu \geq 0$, in which case the Kummer and Tricomi functions reduce to  the well known Hermite polynomials.

   Now consider the effect of the delta function potential. For the odd solutions, the delta function potential is ineffective as the  wave function is zero at the origin.
However, the even parity solutions will be affected, and one requires a discontinuity in the first derivative in the wave function. 
Hence,   in presence of the  delta function  potential, it is the Tricomi function  with non-integer values of $\nu$ which is an appropriate 
solution  having  a discontinuity in the first derivative at the origin. But the Kummer functions are still not acceptable. 
Precisely,  using the Tricomi function, the energy eigenvalues are  obtained  for $g \neq 0$ by  solving the  transcendental 
equation
\be
 \nu - g \frac{\Gamma (1 - \frac{ \nu }{2})}{\Gamma (\frac{1}{2} - \frac{ \nu }{2})} = 0. 
\label{transcendental}
\ee
This equation has to be solved numerically to obtain the energy eigenvalues.

%
\section{Variational method for attractive delta}
\label{sec:attract}

\begin{figure*}
\includegraphics[width=0.4\linewidth]{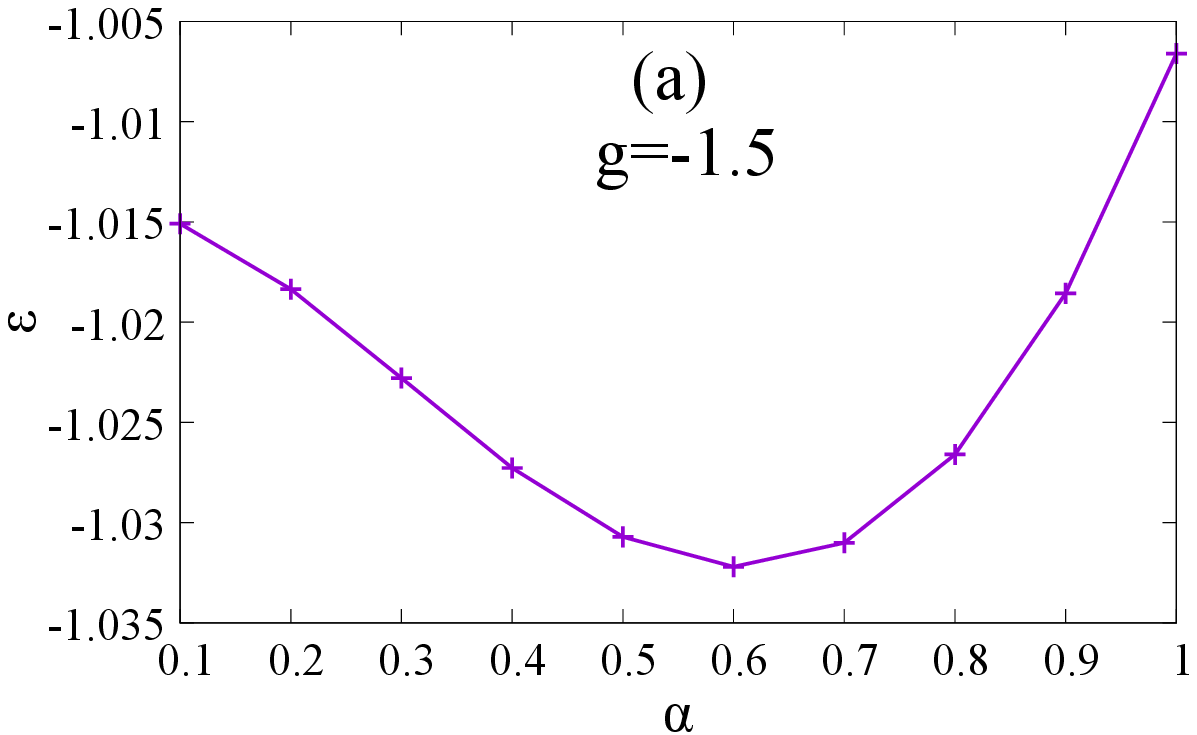}
\includegraphics[width=0.4\linewidth]{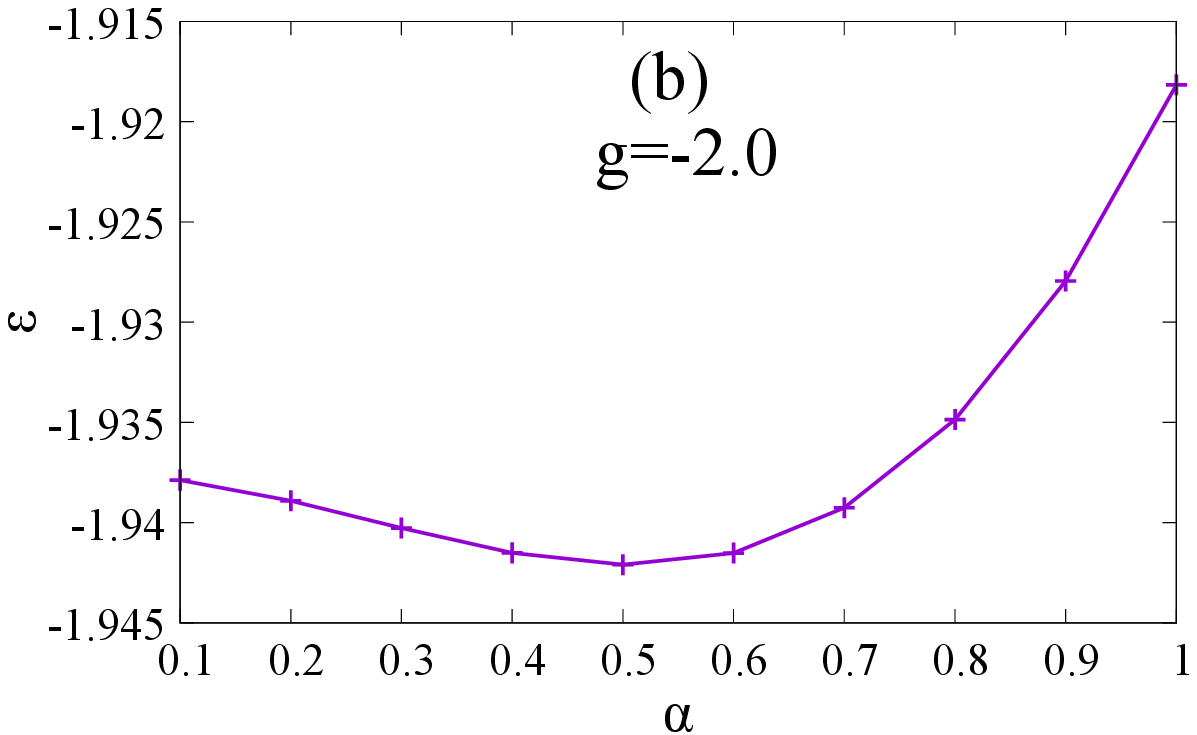}
\caption{ Plots of energy estimates for the attractive delta potential 
as a function of $\alpha$ for $g=-1.5$ and $g=-2.0$ show the approximate location of the minimum value. }
\label{minimum}
\end{figure*}

\begin{figure*}
\includegraphics[width=0.4\linewidth]{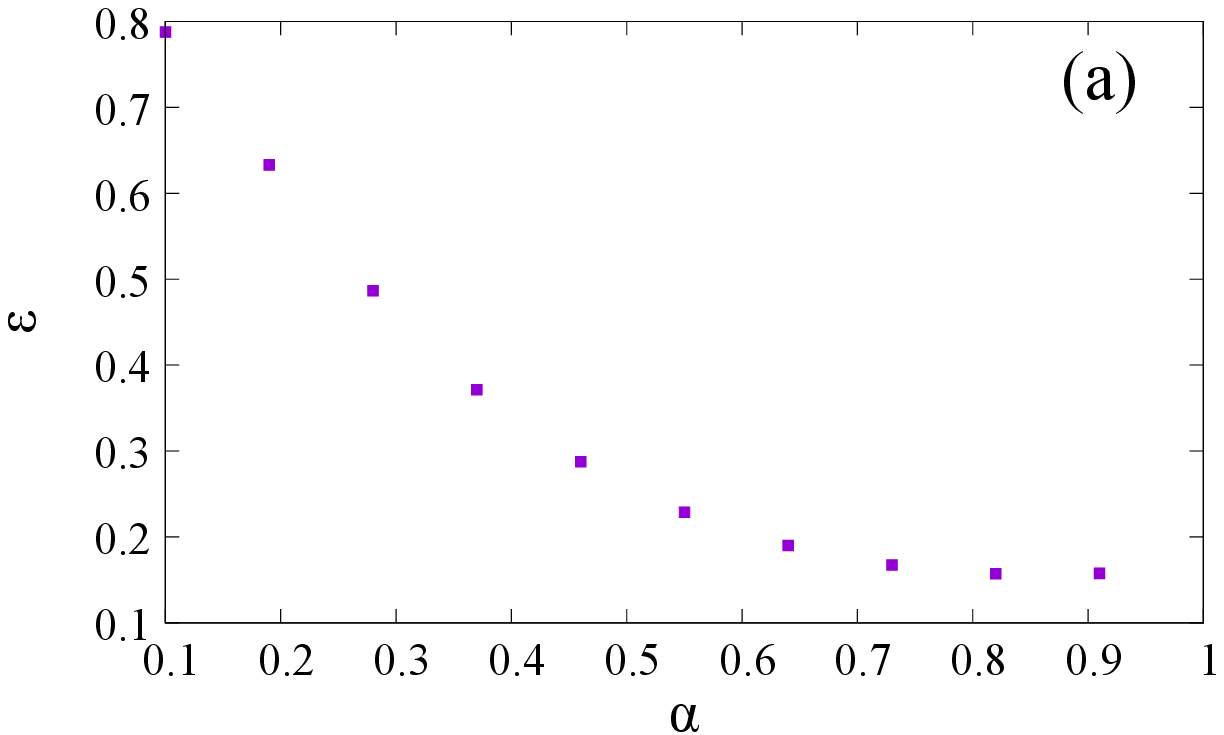}
\includegraphics[width=0.4\linewidth]{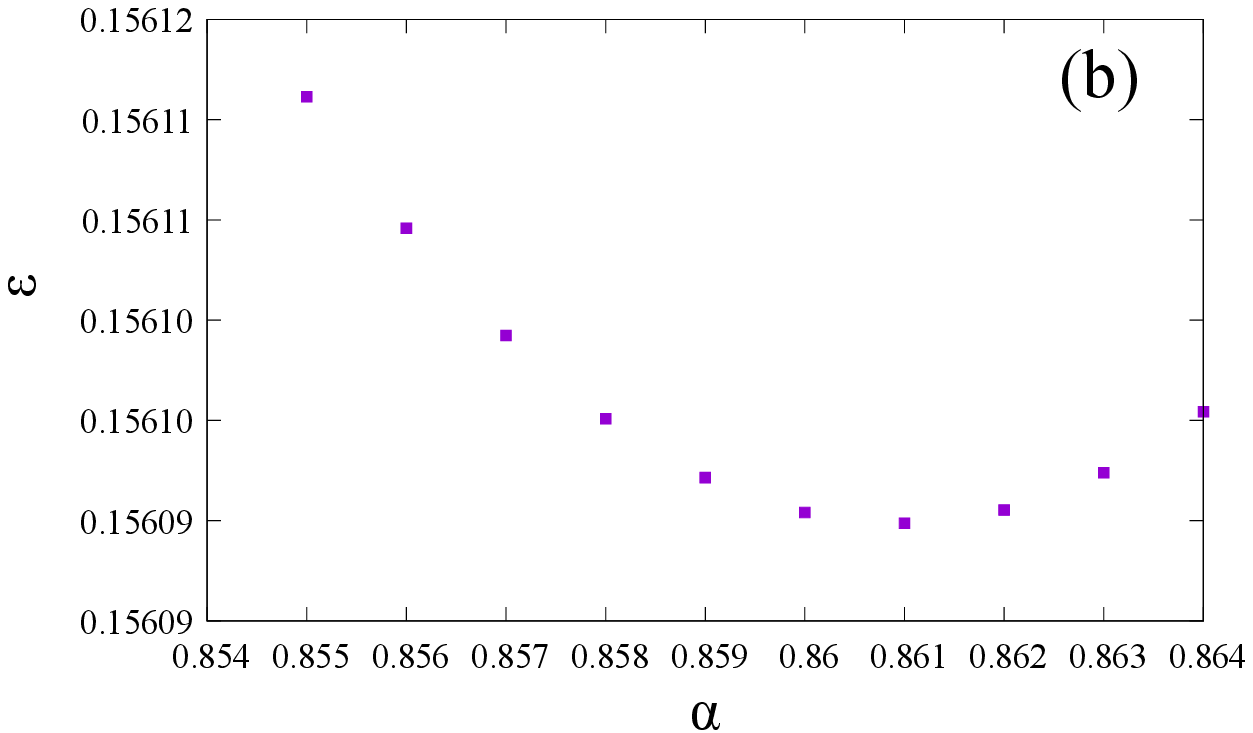}
\includegraphics[width=0.4\linewidth]{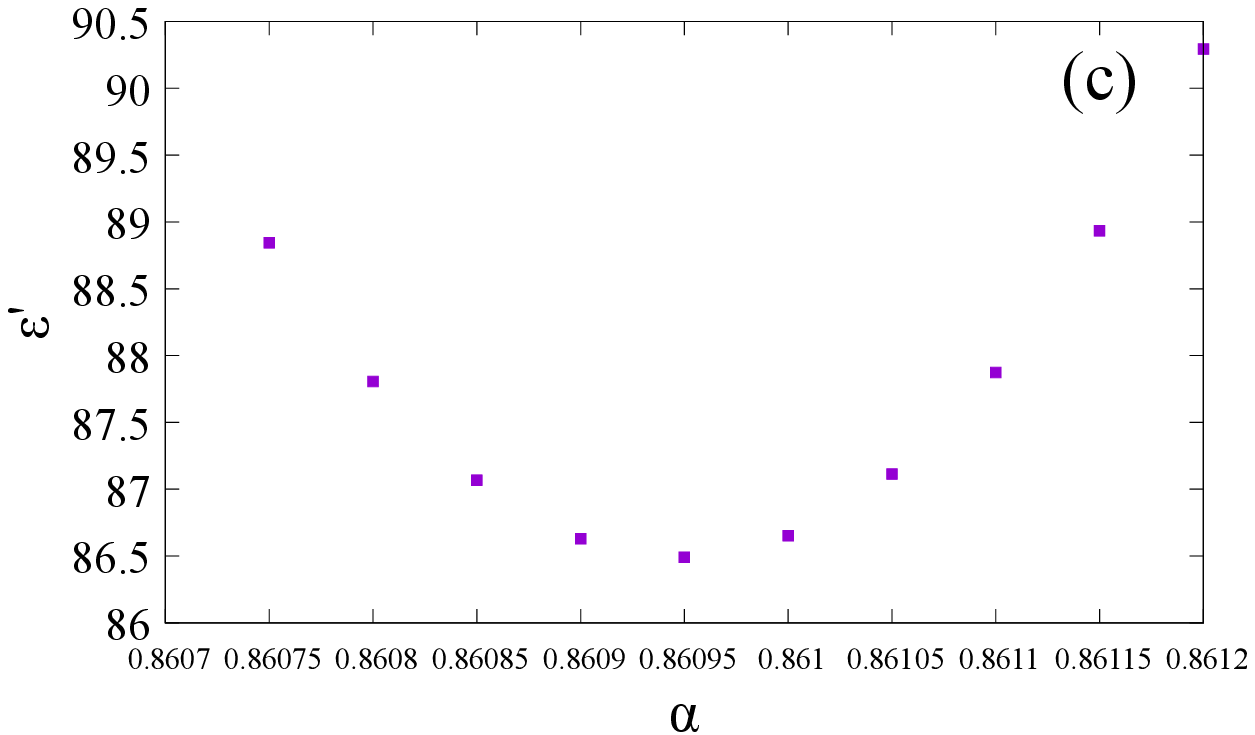}
\includegraphics[width=0.4\linewidth]{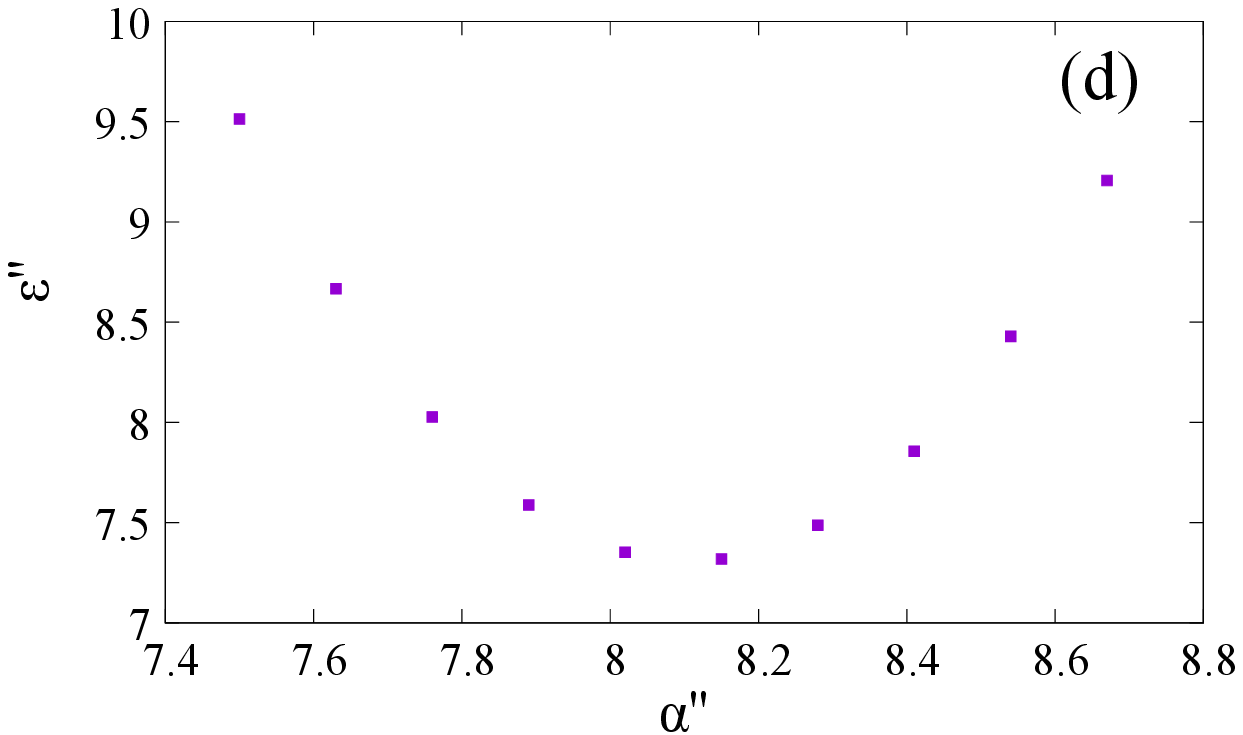}
        \caption{Plots of  $\epsi$ vs $\alpha$ for $g=-0.5$ for some iterative steps  not necessarily in consecutive order. Each iteration zooms in a certain 
smaller interval of $\alpha$ values  studied in the   previous one. Here we show plots 
for intermediate iteration steps when the   size of the interval  is larger than $10^{-7}$.   Table \ref{data} gives the data of the plots. In figure (c) the $\epsi$ values are related to $\epsilon '$  by  $\epsi=0.156089+\epsi '\times 10^{-8}$. In figure (d)  $\alpha$ and $\epsi$ are related to the primed variables as  $\alpha=0.86094+\alpha ''\times 10^{-6}$ and $\epsi=0.156089864909 + \epsi ''\times 10^{-14}$. These plots/data help us to choose the correct interval for further evaluation of $\epsi$ in the next iteration.}
\label{iteration}
\end{figure*}

\begin{table*}
    \centering
    \begin{tabular}{|c|c|c|c|c|c|c|c|}
\hline
\multicolumn{2}{|c|}{Fig \ref{iteration}(a)} & \multicolumn{2}{c|}{Fig \ref{iteration}(b)} & \multicolumn{2}{c|}{Fig \ref{iteration}(c)} & \multicolumn{2}{c|}{Fig \ref{iteration}(d)}\\
\hline
$\alpha$ & $\epsi$ & $\alpha$ & $\epsi$ & $\alpha$ & $\epsi$ & $\alpha$ & $\epsi$\\
    \hline \hline
0.10 & 0.7875843913490 & 0.855 & 0.15611114689438 & 0.86075 & 0.15608988843596 & 0.86094750 & 0.15608986490995\\
0.19 & 0.6330068382654 & 0.856 & 0.15610458365804 & 0.86080 & 0.15608987805867 & 0.86094763 & 0.15608986490987\\
0.28 & 0.4865632769097 & 0.857 & 0.15609922993587 & 0.86085 & 0.15608987067907 & 0.86094776 & 0.15608986490980\\
0.37 & 0.3714877335918 & 0.858 & 0.15609508353865 & 0.86090 & 0.15608986629690 & 0.86094789 & 0.15608986490976\\
0.46 & 0.2874437277571 & 0.859 & 0.15609214228512 & 0.86095 & 0.15608986491188 & 0.86094802 & 0.15608986490974\\
0.55 & 0.2288130659022 & 0.860 & 0.15609040400196 & 0.86100 & 0.15608986652375 & 0.86094815 & 0.15608986490973\\
0.64 & 0.1901599033527 & 0.861 & 0.15608986652375 & 0.86105 & 0.15608987113223 & 0.86094828 & 0.15608986490975\\
0.73 & 0.1672807821995 & 0.862 & 0.15609052769293 & 0.86110 & 0.15608987873706 & 0.86094841 & 0.15608986490979\\
0.82 & 0.1571204709131 & 0.863 & 0.15609238535979 & 0.86115 & 0.15608988933797 & 0.86094854 & 0.15608986490984\\
0.91 & 0.1574914723433 & 0.864 & 0.15609543738243 & 0.86120 & 0.15608990293470 & 0.86094867 & 0.15608986490992\\
\hline
    \end{tabular}
    \caption{Data for the plot in figure \ref{iteration}. From left to right, the values for increasing number of iterative steps are shown. The final value reached with interval size less than $10^{-7}$ is $0.860948$ in this case (not shown in the table). }
    \label{data}
\end{table*}

For the attractive delta potential, $g < 0$. 
We note that any trial  wave function has to satisfy the following criteria\\
(a) It must have definite parity. Only the eigenvalues of even parity solutions will change due to the $\delta$ potential. The eigenvalues of the odd parity states will be identically $(n+\frac{1}{2})\hbar\omega$ ($n=1,3,5, ...$).   \\
(b) It should vanish at infinity\\
(c) The first derivative for the even parity states must have a discontinuity at the origin obeying an equation given later in this section.\\

 In accordance with the above criteria,  we consider trial solutions in the form
\begin{equation}
\label{trial_psi_gen}
\psi (y) = Ae^{Z|y|}e^{-\frac{1}{2}\alpha^2y^{2}}
\end{equation}
for $g < 0$.  Here $A$ is a normalisation constant depending on 
both $Z$ and $\alpha$.

We note that the ground state of the harmonic oscillator should be recovered for $g=0$ such that $Z=0$ and $\alpha =1$ should be the optimal choice. On the other hand, for extremely large values of $g$ one  expects the wave function to be dominantly of the form $\exp (g|y|)$ such that $Z \approx g$ and $\alpha \approx  0$. This indicates that the 
optimal values should follow the bounds: $0 \leq Z \leq g$ and $0 \leq \alpha \leq 1$ with $Z=g$ a sufficient condition at the extreme values. However, the discontinuity condition at the origin, 
\be
 \frac{d\psi}{dy}|_{0^+} - \frac{d\psi}{dy}|_{0^-} = 2g\psi(0) 
\label{disc}
\ee
gives $Z =g$ as a necessary condition when $\psi(y)$ as given in equation \eqref{trial_psi_gen}  is used, which shows that $Z$ cannot be taken as a variable. Here it may be mentioned that the model wave function that was considered in \cite{patil}  had an identical form with fixed values of the parameters; $Z =g$ and $\alpha =1$. In our scheme  we therefore keep only $\alpha$ as   variable  and set $Z=g$ henceforth and obtain the  expectation value of the Hamiltonian $H$ which is expressed in terms of  dimensionless parameter as

\begin{figure*}
\subfigure{\label{comparison_a}\includegraphics[width=0.45\linewidth]{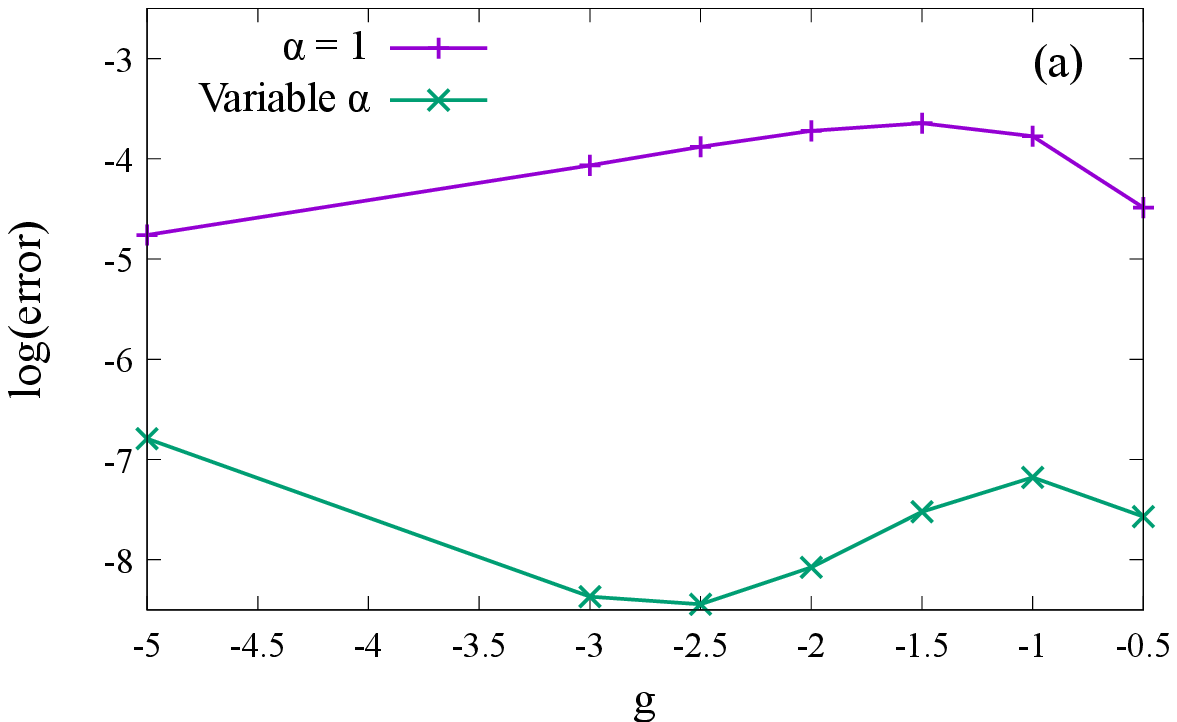}}
\subfigure{\label{comparison_b}\includegraphics[width=0.45\linewidth]{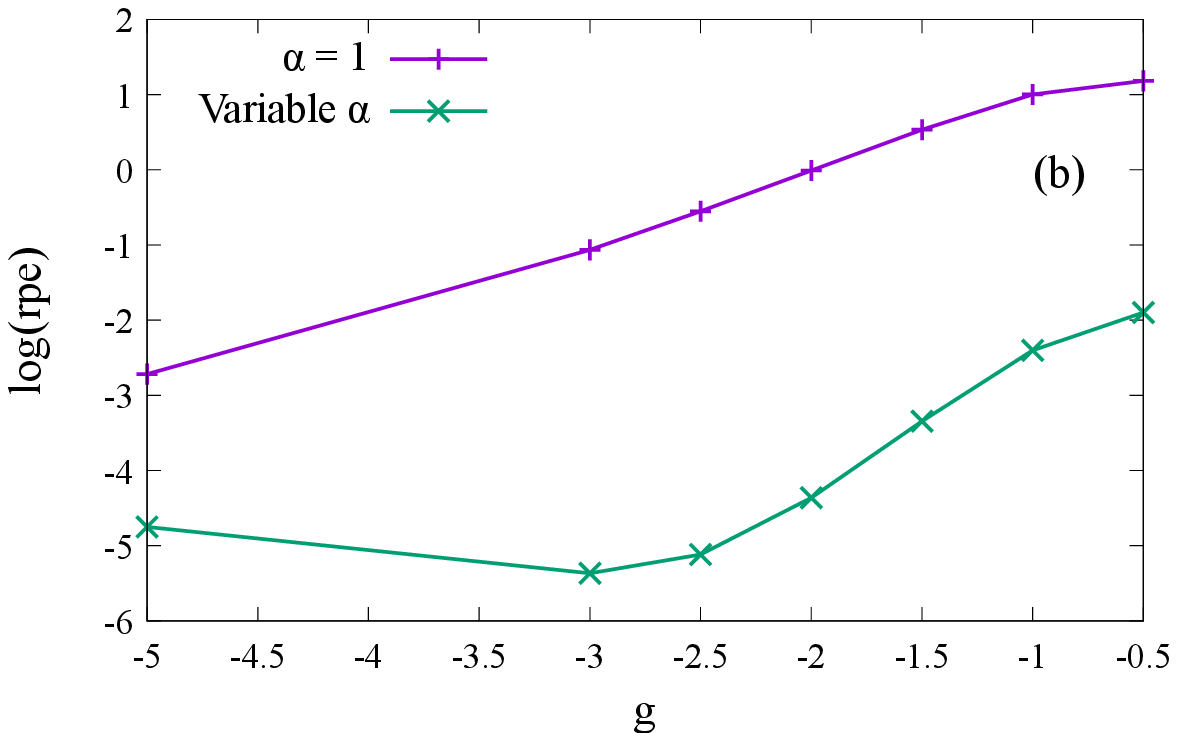}}
\caption{(a) Plots of the error = (energy estimate-exact energy eigenvalue) versus the strength of the attractive delta potential $g$  (b) The relative percentage error ($rpe$) shown against $g$. $rpe$=(error/exact energy value) $\times$  100.
Results for $\alpha=1$ and $\alpha = \alpha_{min}$ are shown for comparison.
}
\label{comparison}
\end{figure*}

\begin{table}
\begin{tabular}{|c|c|c|c|c|c|c|}
\hline
$g \downarrow$&$\alpha_{min}$&$\nu(\alpha_{min})$&$\nu(exact)$&From ref \cite{patil}\\
\hline \hline
-5.0 & 0.219050 & -12.989190 & -12.990313 & -12.981750\\
-3.0 & 0.362841 & -4.972539 & -4.972771 & -4.955630\\
-2.5 & 0.426004 & -3.586291 & -3.5865066 & -3.565851\\
-2.0 & 0.507489 & -2.442049 & -2.442360 & -2.418161\\
-1.5 & 0.609289 & -1.532213 & -1.532729 & -1.506601\\
-1.0 & 0.728909 & -0.841664 &  -0.842418 & -0.819484\\
-0.5 & 0.860948 & -0.343910 & -0.344424 & -0.333176\\
0.1 & 1.023871 &  0.054315 & 0.054269 & 0.054944\\
0.25 & 1.047595  & 0.128397 & 0.128106 & 0.131190\\
0.5 & 1.068158 &  0.234490 & 0.233519 & 0.241000\\
1.0 & 1.077488 &  0.394997 & 0.392743 & 0.404884\\
1.5 & 1.072723 &  0.506696 & 0.503881 & 0.516372\\
2.0 & 1.065157 &  0.586734 & 0.583894 & 0.595116\\
2.5 & 1.057843 &  0.645969 & 0.643356 & 0.652967\\
3.0 & 1.051491 &  0.691160 & 0.688831 & 0.696958\\
5.0 & 1.034671 &  0.797460 & 0.796119 & 0.800388\\
\hline
\end{tabular}
\caption{Summary of results. Values of $\nu$ for different values of $g$ using the approximation method in the present paper are compared with the exact results and previous approximate values,  
re-evaluated up to sixth decimal places. The optimal values of the parameter $\alpha$ are also tabulated. The dimensionless energy $\epsi = (\nu + 1/2)$.}
\label{summary_tab}
\end{table}

\begin{figure}
\includegraphics[width=0.9\linewidth]{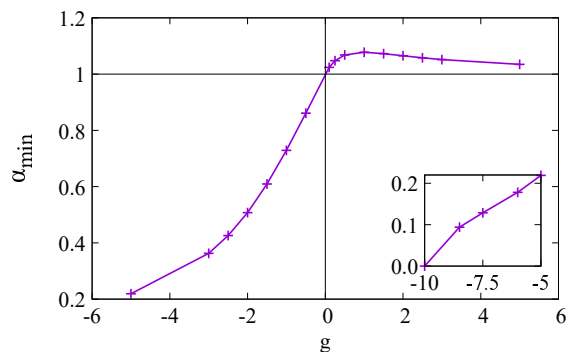}
\caption{The variation of $\alpha_{min}$ with the strength of the potential $g$ is shown. The figure reflects what we expect from physical intuition when the $\delta$ potential is attractive. The inset plot gives the value of $\alpha_{min}$ for large values of $|g|$. This  shows that $\alpha_{min}$ clearly tends to zero
when the delta potential is extremely large in magnitude}.
\label{alphamin_vs_g}
\end{figure}

\begin{equation}
\label{red_eqn}
H=-\frac{1}{2}\frac{d^2}{dy^2}+\frac{1}{2}y^2+g\delta(y). 
\end{equation}
Therefore,  with  $\psi$ given by equation \eqref{trial_psi_gen}, the expectation value of the energy 
$ \epsilon = \la \psi|H |\psi\ra$ is given by  
(see Appendix \ref{appA} for details)  
\begin{gather}
	\epsilon=\frac{\alpha ^2}{2}+\frac{1-\alpha ^4}{2}\frac{\int_{-\infty}^{\infty}y^2e^{2g|y|}e^{-\alpha ^2y^2}dy}{\int_{-\infty}^{\infty}e^{2g|y|}e^{-\alpha ^2y^2} \, dy} \nonumber \\
+ \frac{g^2}{2} + \frac{g}{\int_{-\infty}^{\infty}e^{2g|y|}e^{-\alpha ^2y^2}\, dy} \label{one_param_variation_energy_alpha}
.\end{gather}
 
 After expressing the integrals in terms of complementary error functions ${\text{erfc}} (z)  = \frac{2}{\sqrt{\pi}} \int_z^\infty \exp(-t^2)\,dt$,  we get 

\begin{align}
\label{final_eq}
\epsi = \frac{\alpha ^2}{2}+\frac{1-\alpha^4}{\sqrt{\pi}\alpha^2}\left( \left(\frac{1}{2}+\frac{g^2}{\alpha^2}\right)\frac{\sqrt{\pi}}{2}+\frac{g}{2\alpha}\frac{e^{-\frac{g^2}{\alpha^2}}}{\text{erfc}(-\frac{g}{\alpha})}\right) \nonumber \\
	+\frac{g^2}{2}+\frac{g\alpha~e^{-\frac{g^2}{\alpha^2}}}{\text{erfc}\left(-\frac{g}{\alpha}\right)} 
.\end{align}

The aim is to find the value of $\alpha = \alpha_{min}$ that minimises $\epsi$.    
In principle one can differentiate  $\epsi$ with respect to $\alpha$ and obtain the minimum value using a numerical method  that locates the zero of the derivative.  
However, direct differentiation of $\epsi$ in  equation \eqref{final_eq} to find the optimal value of $\alpha$ leads to a complicated expression involving too many terms. Instead, we use  a  simple indirect  method. 

 We explicitly evaluate  $\epsilon$ for some specific values of  $\alpha$ 
and a  window of $\alpha$ values can be identified within which 
 the minimum is located      
 (indicative  plots are given in figure \ref{minimum} and \ref{iteration} and the data corresponding to one $g$ value   are tabulated  in Table \ref{data}). 
 Further evaluation of $\epsilon$  within this  window   
is made and this process is continued till the desired accuracy up to six decimal places is reached, i.e., when the size of the window is less than $10^{-7}$.  



Both the absolute and relative errors compared to the exact values are shown in figure \ref{comparison} and the results for $\alpha =1$ are also shown for comparison. In Appendix \ref{appB}, the details of calculating numerically the exact values are given. It is  indicated that the present method gives better results compared to the case $\alpha=1$. We have also plotted the optimal values  $\alpha_{min}$ in figure \ref{alphamin_vs_g} and presented the values of $\nu$ and $\alpha$ for some values of $g$ in Table \ref{summary_tab}. The inset of figure \ref{alphamin_vs_g} shows that for large values of the attractive delta potential, $\alpha $ indeed vanishes indicating the harmonic potential has little effect in that limit.
 We will come back to this point in the last section.

\begin{figure*}
\includegraphics[width=0.45\linewidth]{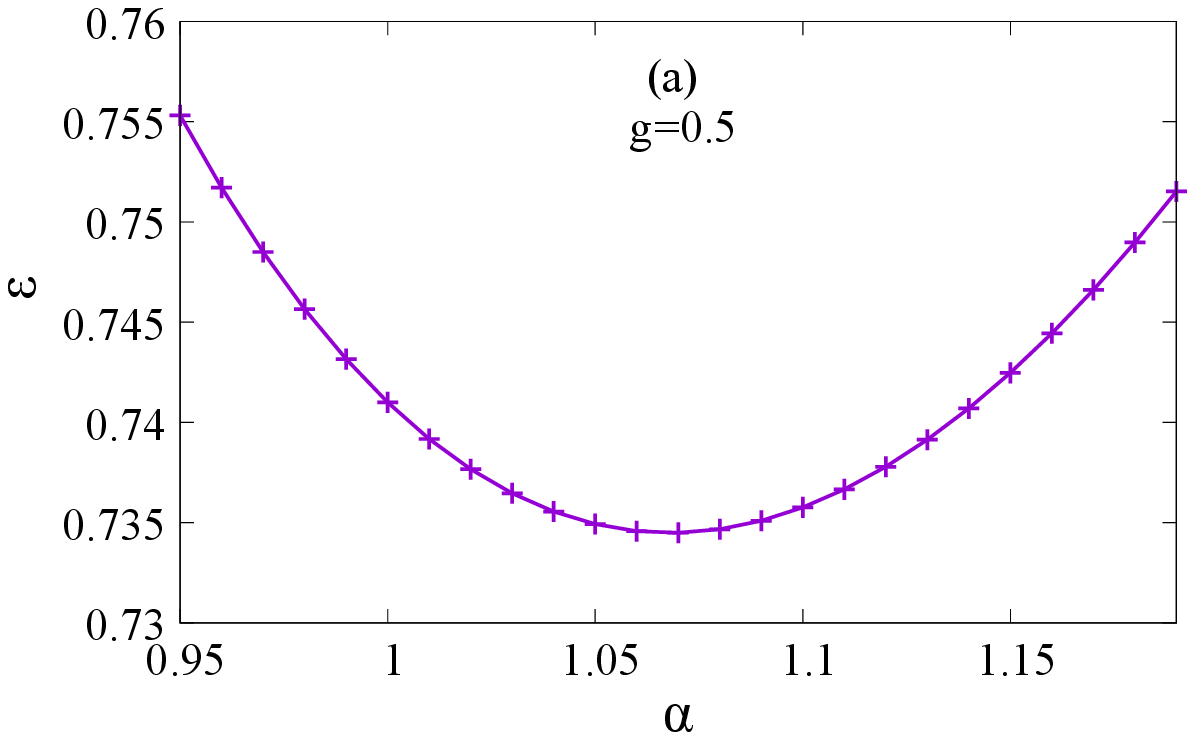}
\includegraphics[width=0.45\linewidth]{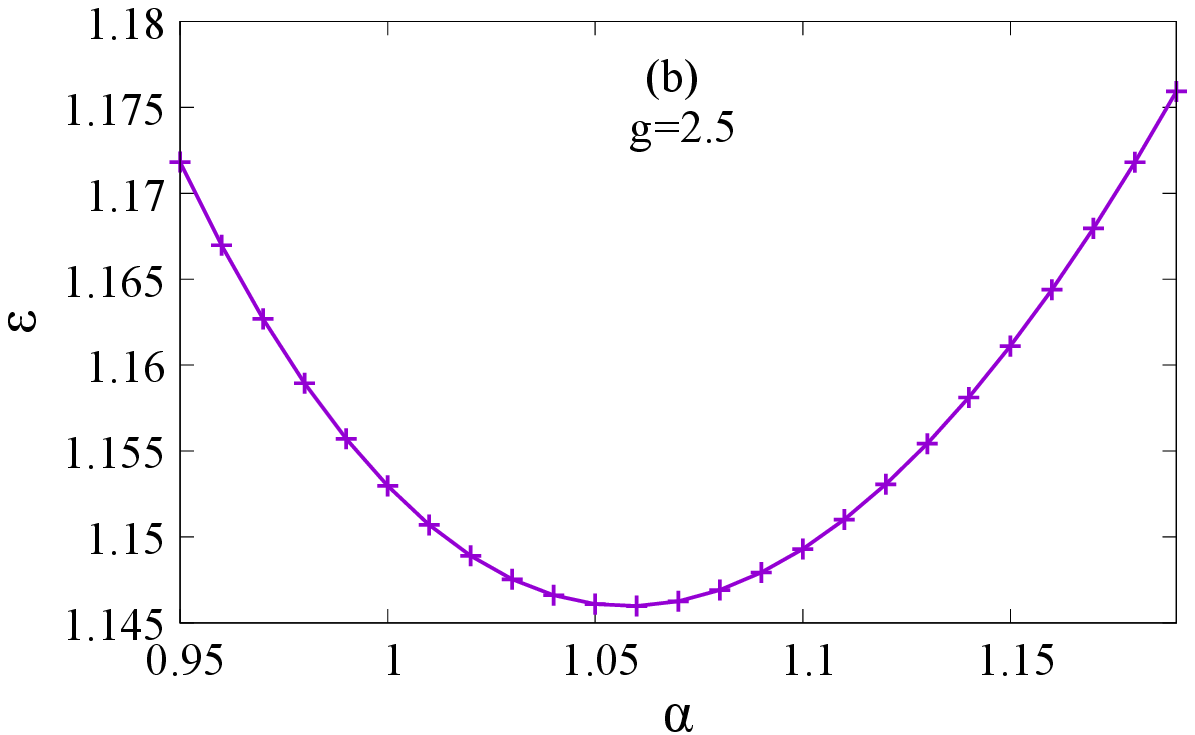}
\caption{Plot of energy estimate as a function of the $\alpha$  for the repulsive delta potential   for $g=0.5$ and $g=2.5$.}
\label{minimum_pos}
\end{figure*}

\begin{figure*}
\includegraphics[width=0.48\linewidth]{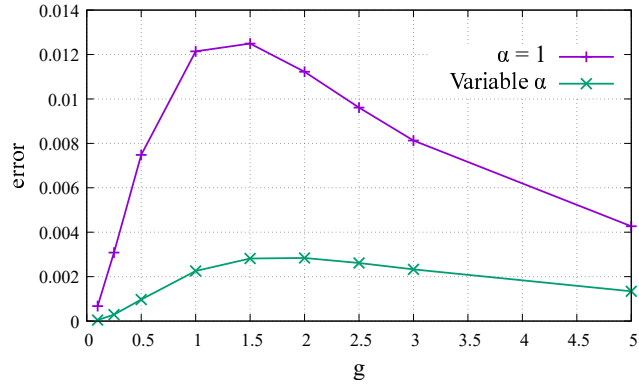}
\caption{Variation of actual error = (energy estimate - exact energy eigenvalue) against the strength of the repulsive delta barrier $g$ for different schemes.}
\label{comparison_pos_g}
\end{figure*}

\section{Variational principle for repulsive delta}
\label{sec:repulse}

The requirements for  the trial wave function for repulsive delta potential are the same as that for the attractive one  mentioned in the beginning of section  \ref{sec:attract}. We use the form of the trial wave function as
\begin{equation}
\label{trial_psi_gen_pos_g}
\psi (y) = A(1+Z|y|)e^{-\frac{1}{2}\alpha ^2y^2}
.\end{equation}

 Once again, this form   (used in  \cite{patil} with $\alpha =1$ and $Z=g$) satisfies the 
discontinuity condition (equation (\ref{disc})) with   $Z=g$ specifically.
In absence of the delta potential, the ground state wave function is therefore recovered and  for very large values of $g$,  $\psi(y) \approx  |y|e^{-\frac{1}{2}\alpha ^2y^2}$, the first excited state of a harmonic oscillator (apart from the absolute value of $y$ which is necessary  for  even parity).   
This is  not surprising as an extremely strong  delta potential will effectively break the system into two adjacent half-harmonic oscillators. We will discuss this in more detail in the next section.

We thus consider the variation of $\epsi$ with respect to $\alpha$ and set $Z=g$ as in the attractive case.
The expectation value of the energy using equation \eqref{trial_psi_gen_pos_g}  is then given by

%
%
\begin{gather}
\label{one_param_variation_energy_alpha_pos_g}
\epsilon=\frac{\alpha^2}{2}+\frac{(1-\alpha^4)}{2}\frac{I}{B}+\frac{g^2}{2}\frac{\frac{\sqrt{\pi}}{\alpha}}{2B}+\frac{g}{2B} \\
\text{where}~I=\frac{\sqrt{\pi}}{4\alpha^3}+\frac{g}{\alpha^4}+\frac{3g^2}{8\alpha^5}\sqrt{\pi} \nonumber \\
B=\frac{\sqrt{\pi}}{2\alpha}+\frac{g}{\alpha^2}+\frac{g^2}{4\alpha^3}\sqrt{\pi} \nonumber
.\end{gather}
Variation of $\epsilon$  against $\alpha$ shows the existence of a minimum as shown in figure \ref{minimum_pos} for  specific values of $g$.
Here it is convenient to directly differentiate $\epsi$ with respect to $\alpha$, and imposing the condition of minimum, we get

\begin{gather}
\frac{\partial \epsilon}{\partial \alpha}=\alpha-\frac{g^2\sqrt{\pi}}{4\alpha^2}\frac{1}{B}-\frac{g^2\sqrt{\pi}}{4\alpha}\frac{1}{B^2}B_{\alpha}-2\alpha^3\frac{I}{B} \nonumber \\
+\frac{(1-\alpha^4)}{2}\frac{I_{\alpha}}{B}-\frac{(1-\alpha^4)}{2}\frac{I}{B^2}B_{\alpha}-\frac{g}{2B^2}B_{\alpha} \label{best_alpha_pos_g} =0\\
\text{where}~I_{\alpha}=\frac{\partial I}{\partial \alpha}=-\frac{3\sqrt{\pi}}{4\alpha^4}-\frac{4g}{\alpha^5}-\frac{15g^2}{8\alpha^6}\sqrt{\pi} \nonumber \\
B_{\alpha}=\frac{\partial B}{\partial \alpha}=-\frac{\sqrt{\pi}}{2\alpha^2}-\frac{2g}{\alpha^3}-\frac{3g^2}{4\alpha^3}\sqrt{\pi} \nonumber
.\end{gather}

The optimal values of $\alpha$ are obtained numerically using bisection method and the results for $\nu$ are presented in Table \ref{summary_tab}  along with the 
exact result and the result from \cite{patil}. Also, we plot the deviations from the exact results in figure \ref{comparison_pos_g} that clearly show that the present results give better approximate values for $\epsi$. The optimal values of $\alpha$ denoted by $\alpha_{min}$ are plotted in figure \ref{alphamin_vs_g}.

\section{Discussions and Comparison with known results}
\label{sec:discuss}

In this paper, we have revisited the problem of the harmonic oscillator with a centrally located $\delta$ potential. Although the results are exactly known, it is useful to apply the variational principle of quantum mechanics to check how close one can approach the exact results and gain insight  from the approximate solutions. We have dealt with the attractive and repulsive delta potential separately, using two different wave functions, taken in the form previously proposed in \cite{patil}. We discuss here the behaviour of the approximate solutions in some limiting cases. In absence of  the delta potential, the ground state wave function of the harmonic oscillator is of the form $e^{-\frac{y^2}{2}}$ which means $\alpha=1$ in equations  \eqref{trial_psi_gen} and \eqref{trial_psi_gen_pos_g}. Hence for $g \to 0 $ we expect $\alpha_{min} \to  1$. This is confirmed from the results for both attractive and repulsive potentials where indeed such a tendency is noted (figure \ref{alphamin_vs_g}).

In presence of an isolated   attractive delta potential (i.e., when there is no harmonic potential), the wave function is of the form $e^{g|y|}$. In the limit $g \to -\infty$ we obtained   $\alpha_{min} \to 0$ using the   variational method 
(figure  \ref{alphamin_vs_g}) indicating  the wave function is 
not affected by the presence of the HO potential.  We present here an argument to 
support this result. 

\begin{figure}
\includegraphics[width=0.9\linewidth]{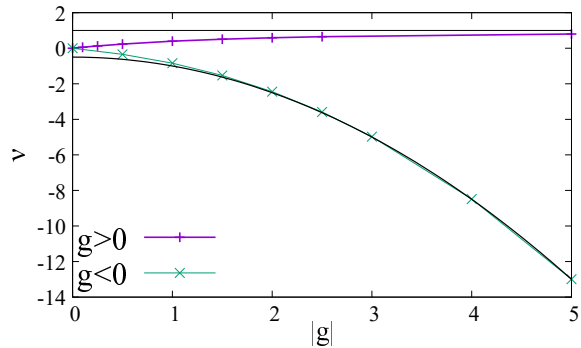}
\caption{The energy quantum number ($\nu$) for some values of strength of the delta potential $|g|$. We also show the curves $-g^2 -1/2$ and $\nu =1$ for comparison.}
\label{nug}
\end{figure}

For the isolated  attractive delta function  there is 
a single  bound state $\psi_0$ with energy  

\begin{equation}
E=-\frac{m\gamma^2}{2\hbar^2} \nonumber \\
\end{equation}
which is negative definite.
Indeed, as $g$ increases in magnitude, this value is approached as shown in the numerical results (see Table \ref{summary_tab};  the data are also shown graphically in figure \ref{nug}) and the corresponding 
 $\nu=\epsilon - \frac{1}{2} = -\frac{g^2}{2}-\frac{1}{2}$.
It is significant that for $g < 0$, $\nu$ has a negative value suggesting 
the delta function potential plays the  dominant role.    
For $g \to -\infty$, we thus argue following  \cite{viana-gomes} that  
the confining harmonic potential, too shallow compared to the 
delta potential,  becomes irrelevant in the ground state. This implies  
  one  can treat the $y^2$ part of the Hamiltonian in equation \eqref{red_eqn} as a perturbation and   $H_0=-\frac{1}{2}\frac{d^2}{dy^2}+g\delta(y)$ is  taken 
as the  unperturbed Hamiltonian. 
The  normalised unperturbed eigenfunction  of $H_0$ is   $\psi_0=\sqrt{|g|}e^{-|g||y|}$. The correction to the unperturbed energy eigenvalue from the first order perturbation theory would then be
\begin{gather}
\braket{\psi_0|\frac{1}{2}y^2|\psi_0}
=|g|\int_{-\infty}^{\infty}e^{-2|g||y|}\frac{1}{2}y^2\,dy=\frac{1}{4g^2}
.\end{gather}
For large values of $|g|$ the correction would therefore tend to vanish and the energy remains as $-g^2/2$.  
Also, there is only one bound state for the pure delta function potential (attractive). Hence,  
the correction to the wave function in the  first order, $\psi^1$,   which has no overlap with this state   is trivially zero. Thus  one would expect $\psi_0$ 
to be the solution even in presence of the HO potential   indicating  
$\alpha \to 0$ in the variational method. 
Note that this argument is valid for the ground state,
for the higher energy states, the confining potential 
will no longer be irrelevant. We also show in figure \ref{nug} that the 
limiting value $\nu= -\frac{g^2}{2}-\frac{1}{2}$ is approached fairly rapidly 
using the variational method.

If $g \to \infty$, for  the repulsive delta potential, we have, as mentioned earlier, effectively two  disconnected half-harmonic oscillators on either side of the origin as the infinite potential 
barrier at the center is effectively insurmountable \cite{aouadi}. In this limit, the   ground state wave function  is given by $A|y|e^{-\frac{1}{2}y^2}$ 
(as $\alpha \to 1$ in  figure \ref{alphamin_vs_g}).  
It is an interesting point that as $g \to \infty$, the value of $\nu$  approaches 1, an  odd integer value  
(as shown in   figure \ref{nug}),
giving rise to the so called anomalous degeneracy. 
 Now we  get two different states belonging to the same eigenvalue $\epsi = \nu +1/2 = 3/2 $ (remember the odd parity state remains unaffected by the delta potential which is characterised 
by an  odd integer value of $\nu $; here we are concerned with the $\nu =1$ state). 
 This result may be apparently  contradictory to the non-degeneracy theorem (and therefore the degeneracy is termed anomalous) in one-dimensional bound states in quantum mechanics. However, exceptions  may occur as the proof of the non-degeneracy theorem rests on a few conditions. If any such condition does not hold good, e.g., if the product of the two supposedly degenerate  wave functions 
is zero, non-degeneracy is not strictly imposed on the spectra of the system \cite{loudon}. 
For the ground state, we have an understanding of the anomalous degeneracy from the present results.  We take the two degenerate solutions as 
$\psi_1 \propto |y|\exp(- y^2/2)$ as given by the trial 
wave function for $g \gg 0$  and $\psi_2 \propto  y\exp(- y^2/2)$, the first order 
excited state (exact). Then $\psi_1\psi_2 =0$ at the origin and $\psi_1 = B \psi_2$ 
albeit with different 
values of $B$ for $x < 0$ and $x > 0$. 
This is consistent with the fact that if the product is zero, one can write $\psi_1 = B \psi_2$ 
 in the regions where the product is non-zero  but 
the constant $B$ may change discontinuously as one passes through a zero of $\psi_1\psi_2$.  
Non-uniqueness in the  value of $B$   implies that the two functions are not trivially    related and degeneracy will then  exist. 
If    $B$ is  different on the two sides of the origin then    
  the slopes of $\psi_1 $ and $\psi_2$ cannot both be continuous at the origin, which is indeed the case here in presence of the delta function. Hence the anomalous degeneracy can exist here.

 To summarise, the behaviour of the system is drastically affected in the   limit
of an  extremely strong delta function, although differently for 
the attractive and repulsive case. In the attractive case, the wave function, 
peaking at the origin does not feel the effect of the harmonic potential ($\propto y^2$) making $\alpha _{min} \to 0$ as $g \to -\infty$ as the factor $exp(g|y|)$ rapidly decreases with the increasing magnitude of $g$.   
For the repulsive case on the other hand, as the wave function peaks away from the origin (occurs at $y \neq 0$), the effect of the harmonic potential is stronger, such that $\alpha_{min}$ is still close to unity.  
In fact $\alpha$ remaining close to unity for the repulsive case for all values of $g$ may be understood using the same logic.

For intermediate values  of $g$ for the attractive delta case, the value of $\alpha_{min}$ obtained from the variational method shows significant deviation from 1, leading to appreciable changes in the energy values 
which are much closer to the exact values. In comparison,   for the repulsive  case,   the energy values are closer to those obtained in \cite{patil} as $\alpha$  remains  fairly close to unity even after applying the variational method.

 One can ask the question,  
why does  the variational method give $\alpha_{min}  \neq 1$? 
The exact solution  is of the form $\exp(-y^2/2)$ multiplied by the Tricomi function.
The trial wave function, on the other hand, has  $\exp(-\alpha^2 y^2/2)$ multiplied by another function 
much  simpler than the  Tricomi function and does not contain $\alpha$, the parameter that is being varied. So the improvement in the result, 
that occurs through varying $\alpha$ must lead to   optimal value of $\alpha \neq 1$ as obtained   
from   the  variational method except at $g=0$.


 That the variational method works  for the higher excited states can also be shown.  We apply it for the known 
first excited state. 
The  trial wave function for the first excited state has to be  orthogonal to the trial wave functions in equation \eqref{trial_psi_gen} and equation \eqref{trial_psi_gen_pos_g}. Since any odd function will be orthogonal to these functions, a viable trial wave function is

\begin{equation}
\label{excited_trial_psi}
\psi(y) = Aye^{Z|y|}e^{-\frac{1}{2}\alpha ^2y^2}
.\end{equation}

The above wave function will do for both $g>0$ and $g<0$. The energy expectation value is

\begin{equation}
\label{variation_first_excited_energy}
\braket{H}=\frac{3\alpha ^2}{2} + \frac{(1 - \alpha ^4)}{2}\frac{\int_0^{\infty} y^4e^{2Zy}e^{-\alpha ^2y^2} \, dy}{\int_0^{\infty} y^2e^{2Zy}e^{-\alpha ^2y^2} \, dy} + \frac{Z^2}{2}.  
\end{equation}

\begin{figure}
\includegraphics[width=8cm]{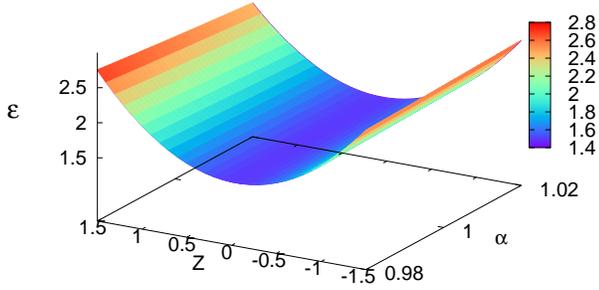}
\caption{Plot of $\epsilon$ from equation \eqref{variation_first_excited_energy} against $\alpha$ and $Z$ showing the optimal value the parameter $Z$ should be zero.}
\label{fig:excited}
\end{figure}

Equation \eqref{variation_first_excited_energy} is independent of $g$ which is physically consistent. So one might as well put $g=0$ which means we have the original HO states intact. This implies that one can put $Z=0$ and $\alpha=1$ in equation \eqref{variation_first_excited_energy}. Even if this is not done directly and  
 both $Z$ and $\alpha$ is varied then from equation \eqref{variation_first_excited_energy}, we get a surface which is shown in Fig. \ref{fig:excited}. As we see the  minima lies at   $Z=0$ 
irrespective of the value of $\alpha$. So, we can put $Z=0$ in equation \eqref{variation_first_excited_energy}. Now the integrals convert into standard Gaussian integrals which can be easily performed. Explicit calculation shows the minimum occurs at $\alpha =1$. Hence one arrives at the exact result. 



  As an endnote, we mention that the  problem of the harmonic oscillator has later been explored extensively introducing more intricacies like multiple delta functions  in arbitrary positions and also in higher dimensions 
\cite{albeverio,aouadi,bush,ferkous,chua}.   
It is also relevant for a charged particle in a magnetic field with a delta potential \cite{olega}. 
The variational method should come in handy when we have to estimate the ground state energy of such cases. As we have shown without doing exact analysis we will be able to calculate accurate values of the energy by trial solutions. It can be a good pedagogical example for students to explore variational principle in a complicated problem and could be simpler in comparison to other 
variational methods like density matrix renormalisation group, also
applied to quantum systems.

This problem has been dealt with in various other contexts also. For example, Bose condensation, which does not occur in one dimension, can happen so in presence of a point interaction localised at the origin \cite{papoyan}. Similarly, the HO in a delta function potential in one dimension could be regarded as a possible model of the three-dimensional hydrogen molecular ion subjected to a static magnetic field in which the coulomb interactions are replaced by the corresponding one of simple point interactions. As the ionisation is relevant only along the direction of the field the three dimensional problem becomes a one dimensional one \cite{dunne,lapidus,lapidus2}. In studying toponium, it was shown  in \cite{avakian} that HO with a point perturbation is a useful model to study quark interactions at short distances. However, such studies have been found to be unphysical later.

Acknowledgements: We are grateful to A, Raychaudhuri for discussions.

ORCID id: Parongama Sen  : 
https://orcid.org/0000-0002-4641-022X

ORCID id: Indrajit Ghose  :
https://orcid.org/0000-0002-8561-4954

\appendix
\section{}
\label{appA}
We give some of the steps to reach equation \eqref{one_param_variation_energy_alpha} from equation \eqref{red_eqn}.

Here 
\be
        \psi=Ae^{g|y|-\frac{1}{2}y^2}  
\ee

Therefore 
\begin{gather}
        \frac{d^2\psi}{dy^2}=\psi(g~\text{sign}(y)-\alpha^2y)^2+\psi(2g\delta(y)-\alpha^2)
\end{gather}

Substituting the expression of $\frac{d^2\psi}{dy^2}$ in equation \eqref{red_eqn} we get a term containing $ye^{2g|y|-\alpha^2y^2}$. We evaluate the term as follows

\begin{gather}
        g\alpha^2\int_{-\infty}^{+\infty}y~\text{sign}(y)|A|^2e^{2g|y|-\alpha^2y^2}\,dy \nonumber \\
        =g\alpha^2e^{2g|y|}[\text{sign(y)}\frac{e^{-\alpha^2y^2}}{-2\alpha^2}]_{-\infty}^{+\infty} \nonumber \\
        +g|A|^2\int_{-\infty}^{+\infty}(g~\text{sign}(y)^2+g\delta(y))e^{2g|y|-\alpha^2y^2}\,dy
\end{gather}

Simple algebra after this will lead to equation \eqref{one_param_variation_energy_alpha}.

\medskip

\section{}
\label{appB}
\renewcommand{\theequation}{\thesection.\arabic{equation}}

In solving the transcendental equation (equation \eqref{transcendental}) we have evaluated the $\Gamma$ by using the defining property of a $\Gamma$ function.

\begin{align}
\Gamma(x)=\frac{\Gamma(u)}{x \times (x-1) \dots \times (u-1)} ~~~ \text{if} ~~~ x<0\\
\Gamma(x)=(x-1) \times (x-2) \dots \times u\Gamma(u) ~~~ \text{if} ~~~ x>0
\end{align}

where $0\leq u \leq 1$ and $\Gamma(u)$ is evaluated by equation \eqref{polynomial_gamma}.

The equation (6.1.35) from \cite{abramowitz} gives
\begin{align}
\Gamma(x+1)=1+b_1x+b_2x^2+b_3x^3+b_4x^4+b_5x^5 \nonumber \\
+b_6x^6+b_7x^7+b_8x^8+\epsilon(x) \label{polynomial_gamma} \\
| \epsilon(x) | \leq 3\times 10^{-7} \nonumber
\end{align}

\begin{align*}
b_1=-&0.577191652&b_5=-&0.756704078\\
b_2=&0.988205891&b_6=&0.482199394\\
b_3=-&0.897056937&b_7=-&0.193527818\\
b_4=&0.918206857&b_8=&0.035868343
\end{align*}

The values of $\nu$ are then calculated numerically using the bisection method.




\end{document}